\documentclass[showkeys,reprint,
superscriptaddress,
 amsmath,amssymb,
 aps,
prb,
floatfix,]{revtex4-1}
\usepackage{graphicx}
\usepackage{dcolumn}
\usepackage{bm}
\usepackage{epstopdf}

\usepackage{amsmath,amssymb,amsfonts}
\usepackage{fancyhdr}
\usepackage{lipsum}
\usepackage{subcaption}
\usepackage{multirow}
\begin{document}

\author{Soumen Mandal}
\email{mandals2@cardiff.ac.uk}
\affiliation{School of Physics and Astronomy, Cardiff University, Cardiff, UK}
\author{Henry A. Bland}
\affiliation{School of Physics and Astronomy, Cardiff University, Cardiff, UK}
\author{Jerome A. Cuenca}
\affiliation{School of Physics and Astronomy, Cardiff University, Cardiff, UK}
\author{Malcolm Snowball}
\affiliation{Ultra Biotecs Limited, Derby, UK}
\author{Oliver A. Williams}
\email{williamso@cardiff.ac.uk}
\affiliation{School of Physics and Astronomy, Cardiff University, Cardiff, UK}

\title{Superconducting boron doped nanocrystalline diamond on boron nitride ceramics}

\begin{abstract}
In this work we have demonstrated the growth of nanocrystalline diamond on boron nitride ceramic. We measured the zeta potential of the ceramics to select the diamond seeds. Diamond was then grown on the seeded ceramics using a microwave chemical vapour deposition system. A clear difference was found between the samples which were seeded with nanodiamond and the ones not seeded before growth. Raman spectroscopy confirmed the excellent quality of the diamond film. Dielectric measurements showed an increase in the dielectric constant of the material after diamond growth. The diamond was also doped with boron to make it superconducting. The film had a transition temperature close to 3.4K. Similar strategies can be applied for growth of diamond on other types of ceramics.
\end{abstract}

\maketitle

\section{Introduction}
Boron and nitrogen belonging to group III and V in the periodic table can combine in a 1:1 ratio to form boron nitride (BN). Primarily it has three different forms, 1. $\alpha$-BN: also known as hexagonal boron nitride (h-BN) or "white graphite" with structure closely matching that of graphite, 2. $\beta$-BN: also known as cubic boron nitride (c-BN) and has structure analogous to diamond, and 3. $\gamma$-BN: which has a wurtzite (w-BN) structure\cite{lipp1989}.  Other forms like turbostratic (t-BN)\cite{tho1962}, rhombohedral (r-BN)\cite{her1958, ish1981} and amorphous\cite{hira1989, ham1993, zed1996} (a-BN) boron nitride are also seen. In this work we will be using ceramics derived from h-BN only. Historically, boron nitride was first synthesised by Balmain in 1842\cite{bal1842} and h-BN was hot pressed into machinable ceramics for the first time in 1952\cite{tay1952}. 

Due to its combination of unique properties like high thermal conductivity, high oxidation resistance, chemical inertness and high dielectric constant it is an excellent candidate for variety of applications\cite{dav1991, hau2002, lee2006, eic2008}. In particular, the dielectric property and high thermal conductivity of boron nitride find usage in electronic and electrical applications\cite{hau2002, wit2015, lat2018}. Further enhancement of dielectric properties in BN ceramics can be achieved by applying a thin coat of high dielectric material like diamond (dielectric constant 5.6\cite{bhag1948}). Since diamond is not machinable as well as forming large work pieces wholly from diamond is extremely expensive it is suitable to coat a thin layer of diamond on machinable ceramics to enhance the property of ceramic. With the advancement of growth technology it is possible to grow diamond on variety of substrates\cite{olie2011}. However, the large difference between the surface energies\cite{har1942, set2002} of diamond and BN makes it impossible to heteroepitaxially grow diamond on BN. Hence, a seeding step is needed to grow diamond on BN ceramics. In the literature, a variety of seeding techniques can be found\cite{olie2011}. For this study we have used an electrostatic nanodiamond colloid based seeding technique. The nanodiamond seeds in the colloid can be both positively or negatively charged depending on the surface termination\cite{hees2011}. So for seeding of the BN substrate it is essential to know the surface charge or zeta ($\zeta$) potential of BN ceramics. In this work we have determined the $\zeta$-potential of the ceramic, which was used to select the diamond seed solution. Diamond thin films were grown on the ceramic using microwave plasma chemical vapour deposition and dielectric constant of the diamond-BN ceramic was measured. Further, to test the properties of diamond film, the layer was doped with boron to test its superconducting properties. The growth of doped diamond on ceramics can be used in areas of electrochemistry, supercapacitors etc where large scale devices can be fabricated with just thin layers of diamond on machinable dielectrics.  Raman studies on doped and undoped diamond layer were done to determine the quality of the layer and scanning electron microscopy was used to observe the crystalline quality and diamond-BN interface of the layer.

\section{Experiment}
The diamond growth was done on commercially available parallel pressed boron nitride ceramics. The binder for the ceramic was boric oxide. The ceramic was shaped into 10 by 10 mm square. The thickness of the squares were 0.5 mm. The dielectric constant of the supplied material at 1 MHz was 4.2 in perpendicular direction as mentioned in the suppliers data sheet. The $\zeta$-potential of BN ceramics was measured using a Surpass$^{TM}$ 3 electrokinetic analyser. The analyser measures the streaming potential by measuring the change in potential or current between two Ag/AgCl electrodes at the ends of a streaming channel as an electrolyte is passed through. Counter-ions from the charged surfaces are sheared by the flowing electrolyte creating a streaming current across the electrodes. The flow of counter-ions is dependent on the electric double layer of the surfaces. Hence, a measure of the streaming current is related to the $\zeta$-potential of the surface\cite{wag1980}.  The setup for measuring zeta potential of flat surfaces was suggested by Van-Wagnen et al.\cite{wag1980}  and has been used successfully to determine the $\zeta$-potential of variety of flat surfaces\cite{voi1983, nor1990, sca1990, man2017}. A 10$^{-3}$ M solution of potassium chloride was used as electrolyte and the pressure was changed between 600 and 200 mbar. The channel width, formed between the BN plates, were kept constant at 100 $\mu$m. The pH of the electrolyte was varied by adding 0.1 M HCl and 0.1 M NaOH solutions with inbuilt titrator in Surpass$^{TM}$ 3. 

\begin{figure}
 \centering
 \includegraphics[height=2.5cm]{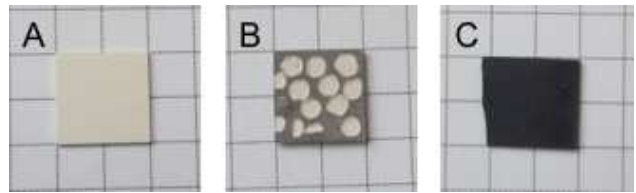}
\caption{Image showing as-received BN ceramic (Panel A). Panel B shows the sample exposed to initial plasma conditions after seeding with nanodiamond but without prebaking. Clear white areas can be seen where the blisters had formed and removed the top surface of the ceramic. Panel C shows a blister free diamond layer on BN ceramics. The sample was seeded with nanodiamond and prebaked before growth.}
\label{sph}
\end{figure}

The diamond films were grown in a Seki Technotron AX6500 series microwave  chemical vapour deposition (CVD) system. A gas mixture of methane, hydrogen and trimethylboron (TMB) was used during the growth of doped diamond. The methane concentration in the mix was 3\%. The TMB was sourced from a premixed hydrogen/TMB gas mixture with 2000ppm TMB concentration. Extra hydrogen was added to the reactor to get the desired methane concentration. The B/C ratio in the gas mixture during growth was calculated to be 12820 ppm. For growing undoped diamond only hydrogen and methane were used. The microwave power and gas pressure during growth were 4 kW and 50 Torr respectively. The temperature of the film during growth, as measured by a Wiliamson dual wavelength pyrometer, was $\sim$800 $^o$C. Before film growth, the substrates were briefly immersed in a mono-dispersed aqueous colloid of hydrogen terminated diamond nanoparticles. Full details for preparation of the diamond seed solution can be found elsewhere\cite{hees2011}. In this case the BN plates were only immersed in the diamond solution not agitated like earlier experiments\cite{klemp2017, klem2017, bla2018}. This is due to the fact that on agitation the BN ceramics started dissociating in the seed solution. After seeding the substrates were rinsed thoroughly in DI water and blow dried. The dried substrates were then prebaked on a hot plate at 120 $^o$C for 12 hours prior to growth. This step was done to expel easily absorbed moisture from the ceramic substrates. The baked substrates were then directly transferred from the hotplate to the reactor to minimise moisture absorption from the atmosphere. In figure \ref{sph} we have shown the image of a bare substrate in Panel A. Panel B shows a seeded substrate, without prebaking, after being exposed to initial plasma conditions. We can clearly see exposed white areas, which are formed due to blister like formation and subsequent peel-off during the initial plasma stage of the growth. The initial plasma condition is when the microwave generator is turned on at low pressures to create the plasma. In our case that is 1.5 kW microwave power and 5 Torr gas pressure. Considering the fact that sample heating in the CVD reactor is due to the plasma, it can be said that the initial plasma conditions will not lead to extreme high temperatures in the substrate. The blistering at low power density conditions may be due to absorbed water in the substrates. The absorbed water interacts with the microwave and is immediately converted to vapour leading to blister formation. To test the hypothesis, we baked a seeded substrate at 120 $^o$C for 12 hours on a hotplate and sample was grown as explained previously. The image of a blister free diamond layer grown by this method is shown in figure \ref{sph}C. From the above it is seems that blister formation is mainly due to absorbed water in the ceramics. Possible ways to get around this problem can be to grow the diamond using a hot filament reactor instead of microwave CVD or replace the hygroscopic boron oxide binder with a non-hygroscopic binder.

Scanning Electron Microscope (SEM) images were taken using a Hitachi SU8200 series FESEM operating at 20 kV and working distances between 9 and 11 mm. Raman measurements were done using an inVia Renishaw confocal microscope using a 514 nm laser.  The excitation wavelength of  514 nm was used as it allows the excitation of both sp$^2$ and sp$^3$ carbon sites in the diamond film\cite{leeds1998}. The dielectric constant and loss tangent of the samples were measured using  a Keysight dielectric test fixture and impedance analyser (16451B and E4099A) in the range of 1 kHz to 1 MHz. More details about the measurement will be published elsewhere\cite{cue2019}. The superconductivity in the films were measured using a Quantum Design Physical Properties Measurement System. The sample was measured in the temperature range of 2-10 K using four silver pasted contacts in the van der Pauw configuration. An excitation current of 5 $\mu$A was used for the measurement. 

\section{Results and Discussion}
\subsection{Zeta Potential}
\begin{figure}[t]
 \centering
 \includegraphics[height=6cm, trim = 90 30 30 30]{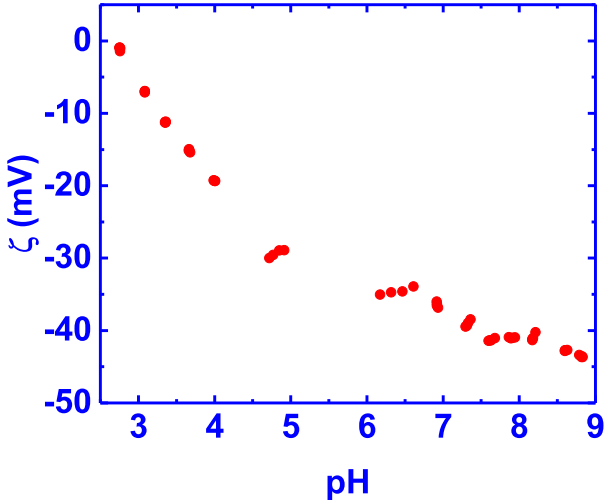}
\caption{Zeta potential of BN ceramics in the pH range 3-9. The zeta potential is mostly negative in the range of measurement. On extending the data the iso-electric point is close to pH 2.6.}
\label{zeta}
\end{figure}
Figure \ref{zeta} shows the $\zeta$-potential of BN ceramic as a function of pH in the range of 3-9. The $\zeta$-potential is negative in the measurement range. On extending the data, we estimate that the isoelectric point (IEP) for the material will be at pH $\sim$ 2.6. Since, the diamond solution used for seeding is in water, the region of our interest is in the pH range 6-7. Considering the negative $\zeta$-potential in this range a hydrogen terminated diamond seed solution, which shows positive $\zeta$-potential\cite{hees2010}, is ideal for seeding the ceramic for diamond growth. Similar negative potential has also been seen for BN solutions\cite{crimp1999, li2015}. In the literature, higher negative $\zeta$-potential in boron ceramics has been attributed to low content of B$_2$O$_3$\cite{crimp1999, dou1991}. The IEP of the ceramic between pH 2-3, points to a strongly acidic solid similar to silica\cite{lewis2004}. This is surprising since boric acid, used as binder for the ceramic, is a weak acid.
\subsection{Scanning Electron Microscopy}
\begin{figure}[!h]
 \centering
 \includegraphics[height=5cm, trim = 30 20 30 30]{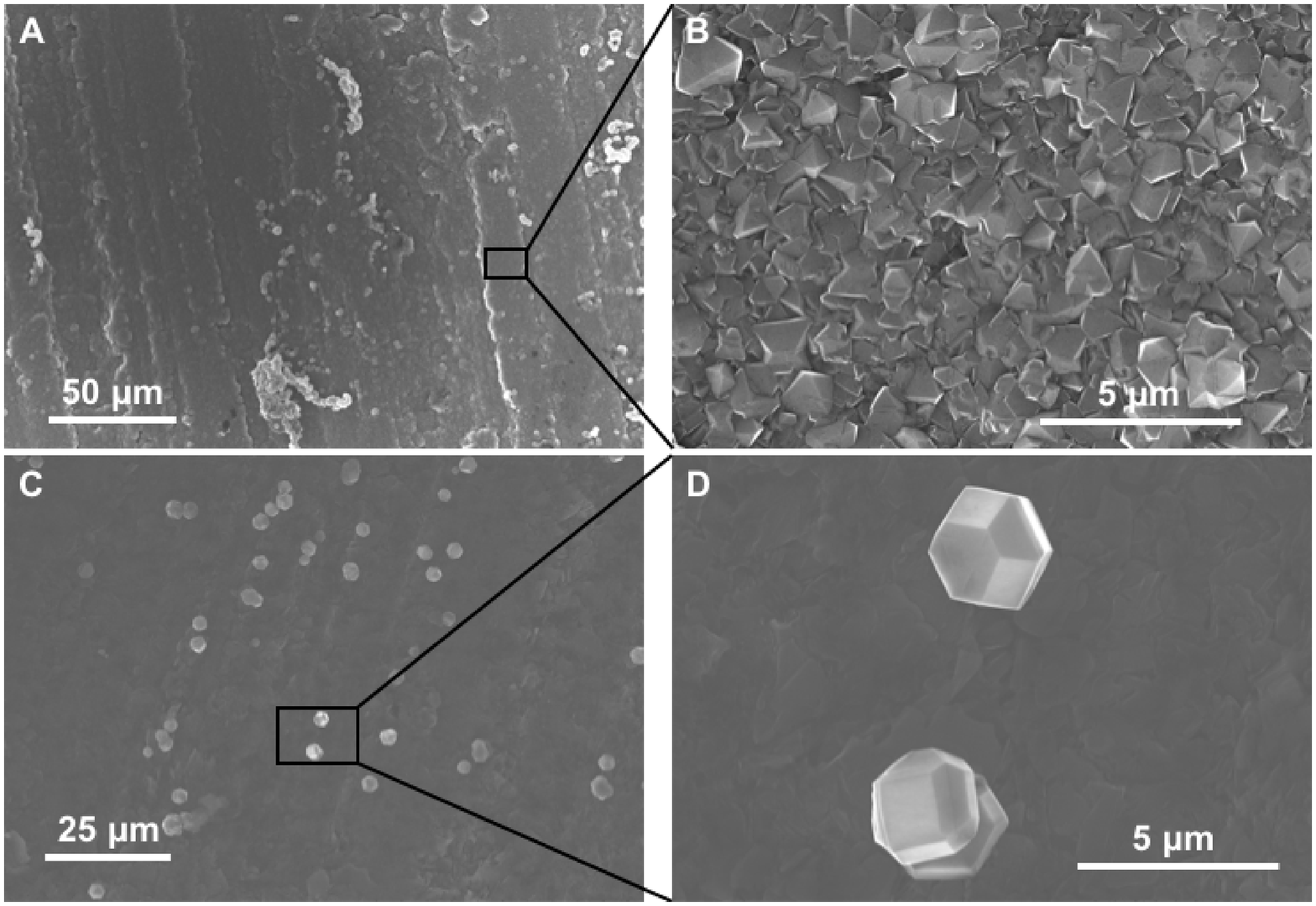}
\caption{SEM images of diamond grown on BN ceramics. Panel A shows a pinhole free diamond film grown on the ceramic seeded with nanodiamond seeds. Panel B is a magnified image of the diamond film shown in panel A. Diamond crystals of varying sizes can be seen which is a characteristic of nanocrystalline diamond films. Panel C shows growth of individual seeds on the composite. The composite was unseeded to study the effectiveness of seeding. Panel D shows magnified image of two micro crystals grown on unseeded BN composite.}
\label{sus}
\end{figure}

\begin{figure}
 \centering
 \includegraphics[height=2.5cm, trim = 30 20 30 15]{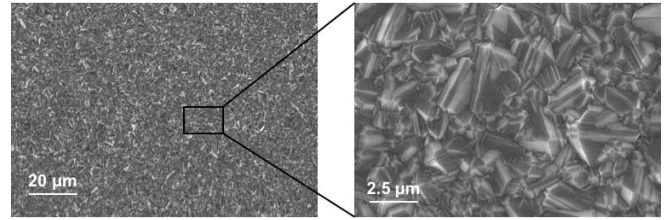}
\caption{SEM images of diamond grown on BN ceramics which were seeded by brushing dry nanodiamond powder. Panel A shows a pinhole free diamond layer. Panel B shows a magnified image of the diamond layer showing nanodiamond crystals.}
\label{ps}
\end{figure}

Figure \ref{sus} shows the SEM images of diamond grown on BN ceramics. As evident from figure \ref{zeta}, we have to use positively charged diamond particles for seeding. The seeding was done using a nanodiamond solution containing 5nm hydrogen terminated nanodiamond particles. To check the effectiveness of the seeding process we have grown diamond on both seeded and unseeded substrate in the same growth run. Figure \ref{sus}A shows the diamond grown on a seeded BN composite while panel C in the same figure shows the growth on unseeded substrate. The seeded substrate shows a pinhole free diamond film. A magnified view of the diamond film is shown in figure \ref{sus}B. The growth of diamond on unseeded BN composite, shown in figure \ref{sus}C, is sparse and we see individual grains on the surface. A magnified view of the individual diamond grains show good quality diamond grains (figure \ref{sus}D). Also, we have tried growing diamond on composites brushed with 5 nm diamond seeds. The SEM images of such films is shown in figure \ref{ps}. In panel A of figure \ref{ps} we have shown the overview of the diamond with pinhole free growth.  A magnified view of the film with good quality crystal is shown in figure \ref{ps}B. Even though at the microscopic scale there is not much difference between films grown by colloid or brushing the nanodiamond on composite, using the colloid enabled growth of blister free samples. Such blister free samples were not possible even after post baking with brushing the nanodiamond seeds (see figure \ref{sph}B). Lastly, we have grown a sample with brushing of nanodiamond seeds after wetting the substrate in DI water for 10 minutes. In this case as well blister free samples could be grown. This means wetting and postbaking is important for blister free growth using microwave CVD.

\begin{figure}[!h]
 \centering
 \includegraphics[height=15cm]{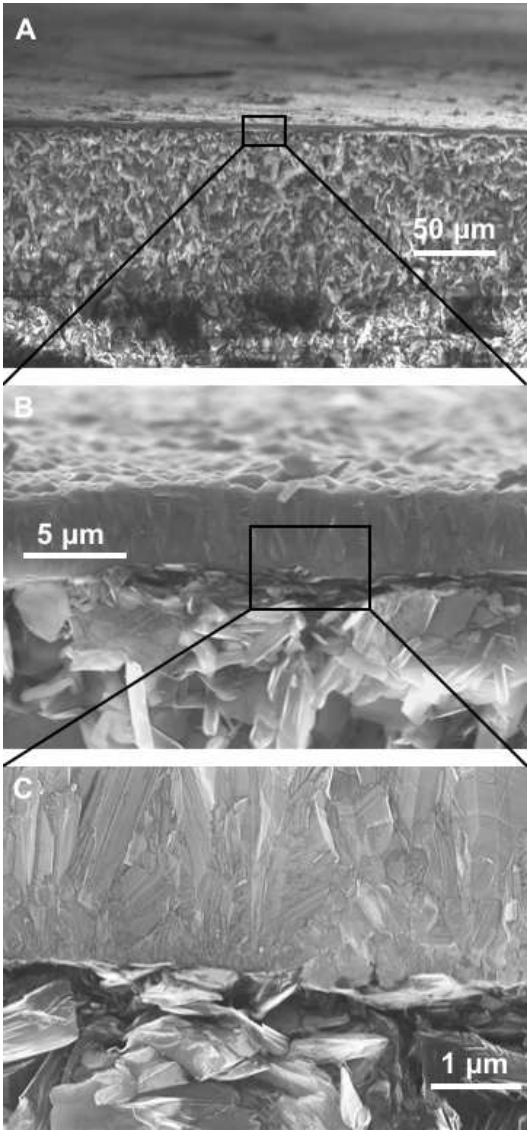}
\caption{SEM image of cross-section of diamond grown on BN ceramics. Panel A shows an overview of the diamond grown on BN ceramics. We can see the thin diamond layer on the top. Panel B and Panel C are successive magnified images of the interface.}
\label{scs}
\end{figure}

Finally, we have imaged the cross-section of the diamond-BN composite. Figure \ref{scs} shows the cross-section of the diamond-BN interface. In panel A we have shown the overview of the film grown on BN. Panel B and C shows the subsequent magnified image of the interface as specified in figure by the black squares. In panel B we can clearly see the dense diamond layer on top. The BN composite with elongated nitride crystals can be seen at the bottom. The diamond layer is approximately 5 micron thick in this case. The magnified image of the interface in panel C show a continuous diamond layer without large scale voids. The dense diamond layer can be seen growing well on top of the pressed ceramic.


\subsection{Raman Spectroscopy}
\begin{figure}[!h]
 \centering
 \includegraphics[height=6cm, trim = 90 80 90 50]{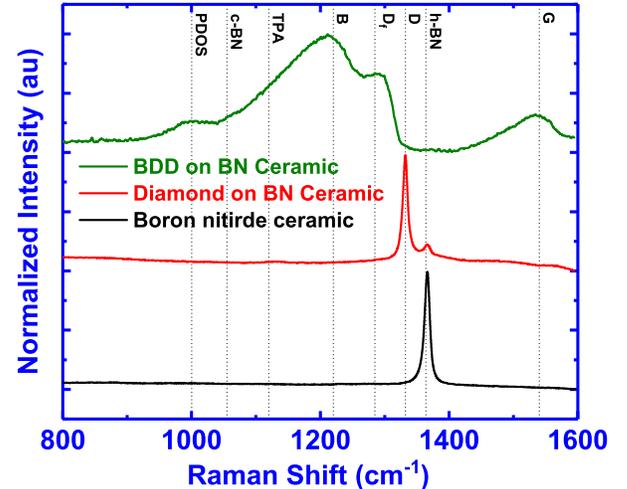}
\caption{Raman spectra of bare substrate(black), intrinsic nanocrystalline diamond(red) and boron doped nanocrystalline diamond(green) grown on BN ceramics. The dotted lines show the positions of various standard peaks.}
\label{raman}
\end{figure}

In this work we have grown both doped and undoped diamond on BN ceramics.  Figure \ref{raman} shows the Raman spectra of the bare substrate (black), intrinsic nanocrystalline diamond (red) and boron doped nanocrystalline diamond (green) grown on BN substrate. The dotted lines show the position of standard peaks. The h-BN and c-BN peaks should appear at 1364 and 1055 cm$^{-1}$ respectively\cite{reich2005}. The bare substrate shows a sharp h-BN peak and no c-BN peak indicating the presence of only h-BN in the composite. The spectra of undoped diamond grown on the composite is shown in red. We see a clear diamond peak\cite{ram1930, robert1930, bhag1930}  at 1332 cm$^{-1}$ along with a smaller h-BN peak. The diamond layer does not show any non-diamond carbon in the layer. Finally we have done Raman on the boron doped diamond (BDD) layer grown on the composite. The spectra is similar to BDD layers grown on silicon substrates\cite{may2008, szi2012}.

The film shows the regular characteristics of a heavily boron doped diamond with a broad band, B, peak at 1220 cm$^{-1}$ and a shoulder D$_F$ at 1285 cm$^{-1}$\cite{szi2012, bla2018}. Sidorov and Ekimov\cite{sidorov2010} attribute the 1220 cm$^{-1}$ peak to carbon-carbon bonding states. The addition of boron to diamond leads to locally distorted lattice structure which can give rise to the band like structure\cite{ash2013} and the intensity of the band is directly related to the doping level\cite{may2007}.  The 1285 cm$^{-1}$ shoulder is the red shifted diamond line from the characteristic 1332 cm$^{-1}$ line seen in pure diamond. The Fano-like line shape of the diamond line is attributed to the quantum interference between the zone centre diamond phonon and the electronic states continuum introduced by the dopant\cite{ghee1993, ager1995, pru2001}. The trans-polyacetylene (TPA) shoulder at 1120 cm$^{-1}$ can be clearly seen in the BDD film while that is absent in the undoped diamond layer. TPA is generally found at the grain boundaries in nanocrystalline diamond (NCD)\cite{fer2001}. Furthermore, we can see a clear G band at 1560 cm$^{-1}$ in the case of BDD while this is not present for undoped diamond. The G peak is due to the in-plane stretching of a pair of sp$^2$ carbon sites\cite{fer2000, fer2001a, fer2004}. Commonly the G band is seen at 1580 cm$^{-1}$ but it can appear anywhere between 1520-1580 cm$^{-1}$\cite{pra2004}. The shift in the G-band from 1580 cm$^{-1}$ is mainly due to a sp$^2$ site converting to sp$^3$ site when a $\pi$ ring system converts to a $\pi$ chain system\cite{fer2004}. The absence of the G band in undoped diamond implies the high quality of the NCD film on BN ceramic. Finally, a small shoulder at 1000 cm$^{-1}$ can be seen which has been attributed to the maximum of phonon density of states (PDOS) in diamond\cite{szi2012}. This occurs due to the defects introduced by the dopant, allowing for the forbidden states to appear\cite{pra1998}.
\subsection{Dielectric Constant}
We have measured the dielectric constant of the as received BN ceramics and the ceramics with undoped diamond grown on it. The measured dielectric constant of the pure boron nitride is frequency independent, with a nominal value of 3.9 $\pm$ 0.1 across the range of measurement. In addition, owing to its insulating nature, the loss tangent is immeasurable with the measurement setup. The dielectric constant for the diamond-BN hybrid, measured through the material, increases to 9.5 $\pm$ 0.3. The dielectric loss tangent also increases to approximately 0.01 $\pm$ 0.002. For ceramic materials the dominant polarisation mechanism that contributes to the dielectric constant and loss, otherwise known as the complex permittivity, is electronic polarisation. This mechanism is associated with the displacement of the electron cloud distribution with respect to the nucleus which resonates out in the high terahertz region. This explains why the complex permittivity of BN ceramic is  frequency independent at kilohertz frequencies. Additionally, BN is not polar nor is it ionic since the B and N atoms share the same number of electrons, therefore, dipolar relaxation and ionic polarisation are not present. The immeasurable dielectric loss tangent is congruent with the insulating nature of h-BN owing to its high band gap energy. Even though sp$^2$ bonded, the electrical properties of h-BN are unlike the carbon equivalent of graphite, where a high dielectric loss tangent is expected owing to its high planar conductivity. After CVD diamond growth, there is a large increase in dielectric constant. While the dielectric constant of diamond is approximately 5.6\cite{bhag1948}, the increase may be associated with Maxwell-Wagner-Sillars type interfacial polarisation\cite{xia2017}. This model is normally used for explaining electrically conducting materials in a non-conducting medium; both h-BN and diamond are electrical insulators. In which case, very small concentrations of sp$^2$ carbon may be present at the interface between the materials and therefore, the absence of sp$^2$ carbon signatures in the Raman spectroscopy measurements is plausible\cite{cue2018}. 

\subsection{Superconductivity}
\begin{figure}[!h]
 \centering
 \includegraphics[height=6cm, trim = 90 30 30 30]{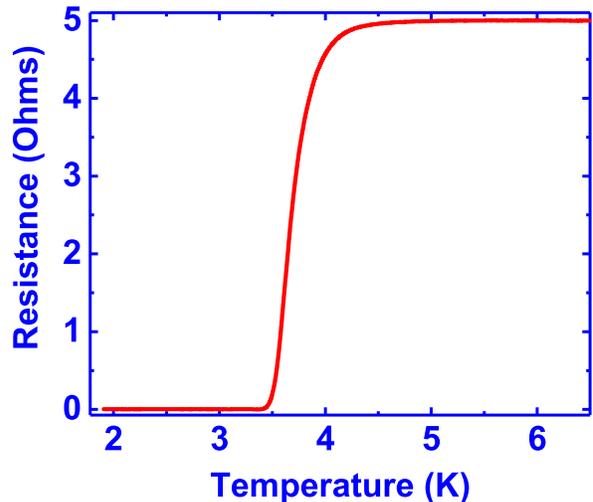}
\caption{Resistance vs temperature of BDD grown on BN ceramics. The film shows a superconducting transition temperature at 3.5K.}
\label{rt}
\end{figure}
Superconductivity in boron doped diamond\cite{ekimov2004} is one of the many superlative properties of BDD. We have tested the superconducting properties of doped diamond on BN ceramics to see if this property is retained even when the material is grown on a ceramic.  Figure \ref{rt} shows the resistance vs temperature for a BDD film grown on BN ceramics. The transition temperature (T$_c$) of the film is $\sim$ 3.4K and the transition width, $\Delta T_c$ = 1K. As the film was grown on an insulating ceramic like BN, the superconductivity is primarily from the BDD layer. The transition temperature was determined as the temperature at which the resistance is 1\% of the normal state resistance. The transition width is the temperature difference between the temperatures at which the sample resistance is 99\% and 1\% of the normal state resistance\cite{bla2018}.  Similar transition temperatures and width have been reported on BDD\cite{ekimov2004, takano2005, klem2017, zhang2017}. From figure \ref{rt} it is clear that the superconductivity in BDD is unaffected by the substrate material. This technique of growing diamond on ceramics can be useful for electrochemistry, supercapacitors etc. where machinable dielectrics can form large scale devices, which can then be coated with a thin layer of BDD.

\section{Conclusion}
In conclusion, we have determined the $\zeta$-potential of BN-ceramics. The $\zeta$-potential is negative in the pH range of the diamond colloids used for seeding. The growth of diamond on the ceramic involves a pre-baking step due to the hygroscopic nature of the ceramic binding material. The quality of diamond as determined by Raman spectroscopy and SEM was found to be excellent. An increase in dielectric constant from  3.9 $\pm$ 0.1 to 9.5 $\pm$ 0.3 was observed. Doped diamond grown on the ceramic retained its excellent superconducting property.

\section*{Acknowledgment}
SM and OAW acknowledge financial support of the European Research Council (ERC) Consolidator Grant SUPERNEMS, Project ID: 647471. JAC acknowledge financial support of the Engineering and Physical Sciences Research Council under the program Grant GaN-DaME (EP/P00945X/1). 

\section*{Dataset}
The datasets generated and/or analysed during this study can be found at http://doi.org/10.17035/d.2019.0069330616

\bibliography{rsc}

\begin{thebibliography}{58}%
\makeatletter
\providecommand \@ifxundefined [1]{%
 \@ifx{#1\undefined}
}%
\providecommand \@ifnum [1]{%
 \ifnum #1\expandafter \@firstoftwo
 \else \expandafter \@secondoftwo
 \fi
}%
\providecommand \@ifx [1]{%
 \ifx #1\expandafter \@firstoftwo
 \else \expandafter \@secondoftwo
 \fi
}%
\providecommand \natexlab [1]{#1}%
\providecommand \enquote  [1]{``#1''}%
\providecommand \bibnamefont  [1]{#1}%
\providecommand \bibfnamefont [1]{#1}%
\providecommand \citenamefont [1]{#1}%
\providecommand \href@noop [0]{\@secondoftwo}%
\providecommand \href [0]{\begingroup \@sanitize@url \@href}%
\providecommand \@href[1]{\@@startlink{#1}\@@href}%
\providecommand \@@href[1]{\endgroup#1\@@endlink}%
\providecommand \@sanitize@url [0]{\catcode `\\12\catcode `\$12\catcode
  `\&12\catcode `\#12\catcode `\^12\catcode `\_12\catcode `\%12\relax}%
\providecommand \@@startlink[1]{}%
\providecommand \@@endlink[0]{}%
\providecommand \url  [0]{\begingroup\@sanitize@url \@url }%
\providecommand \@url [1]{\endgroup\@href {#1}{\urlprefix }}%
\providecommand \urlprefix  [0]{URL }%
\providecommand \Eprint [0]{\href }%
\providecommand \doibase [0]{http://dx.doi.org/}%
\providecommand \selectlanguage [0]{\@gobble}%
\providecommand \bibinfo  [0]{\@secondoftwo}%
\providecommand \bibfield  [0]{\@secondoftwo}%
\providecommand \translation [1]{[#1]}%
\providecommand \BibitemOpen [0]{}%
\providecommand \bibitemStop [0]{}%
\providecommand \bibitemNoStop [0]{.\EOS\space}%
\providecommand \EOS [0]{\spacefactor3000\relax}%
\providecommand \BibitemShut  [1]{\csname bibitem#1\endcsname}%
\let\auto@bib@innerbib\@empty
\bibitem [{\citenamefont {Lipp}\ \emph {et~al.}(1989)\citenamefont {Lipp},
  \citenamefont {Schwetz},\ and\ \citenamefont {Hunold}}]{lipp1989}%
  \BibitemOpen
  \bibfield  {author} {\bibinfo {author} {\bibfnamefont {A.}~\bibnamefont
  {Lipp}}, \bibinfo {author} {\bibfnamefont {K.}~\bibnamefont {Schwetz}}, \
  and\ \bibinfo {author} {\bibfnamefont {K.}~\bibnamefont {Hunold}},\ }\href
  {\doibase 10.1016/0955-2219(89)90003-4} {\bibfield  {journal} {\bibinfo
  {journal} {Journal of the European Ceramic Society}\ }\textbf {\bibinfo
  {volume} {5}},\ \bibinfo {pages} {3} (\bibinfo {year} {1989})}\BibitemShut
  {NoStop}%
\bibitem [{\citenamefont {Thomas}\ \emph {et~al.}(1962)\citenamefont {Thomas},
  \citenamefont {Weston},\ and\ \citenamefont {O'Connor}}]{tho1962}%
  \BibitemOpen
  \bibfield  {author} {\bibinfo {author} {\bibfnamefont {J.}~\bibnamefont
  {Thomas}}, \bibinfo {author} {\bibfnamefont {N.~E.}\ \bibnamefont {Weston}},
  \ and\ \bibinfo {author} {\bibfnamefont {T.~E.}\ \bibnamefont {O'Connor}},\
  }\href {\doibase 10.1021/ja00883a001} {\bibfield  {journal} {\bibinfo
  {journal} {Journal of the American Chemical Society}\ }\textbf {\bibinfo
  {volume} {84}},\ \bibinfo {pages} {4619} (\bibinfo {year}
  {1962})}\BibitemShut {NoStop}%
\bibitem [{\citenamefont {H{\'{e}}rold}\ \emph {et~al.}(1958)\citenamefont
  {H{\'{e}}rold}, \citenamefont {Marzluf},\ and\ \citenamefont
  {P{\'{e}}rio}}]{her1958}%
  \BibitemOpen
  \bibfield  {author} {\bibinfo {author} {\bibfnamefont {A.}~\bibnamefont
  {H{\'{e}}rold}}, \bibinfo {author} {\bibfnamefont {B.}~\bibnamefont
  {Marzluf}}, \ and\ \bibinfo {author} {\bibfnamefont {P.}~\bibnamefont
  {P{\'{e}}rio}},\ }\href@noop {} {\bibfield  {journal} {\bibinfo  {journal}
  {Comptes Rendus}\ }\textbf {\bibinfo {volume} {246}},\ \bibinfo {pages}
  {1866} (\bibinfo {year} {1958})}\BibitemShut {NoStop}%
\bibitem [{\citenamefont {Ishii}\ \emph {et~al.}(1981)\citenamefont {Ishii},
  \citenamefont {Sato}, \citenamefont {Sekikawa},\ and\ \citenamefont
  {Iwata}}]{ish1981}%
  \BibitemOpen
  \bibfield  {author} {\bibinfo {author} {\bibfnamefont {T.}~\bibnamefont
  {Ishii}}, \bibinfo {author} {\bibfnamefont {T.}~\bibnamefont {Sato}},
  \bibinfo {author} {\bibfnamefont {Y.}~\bibnamefont {Sekikawa}}, \ and\
  \bibinfo {author} {\bibfnamefont {M.}~\bibnamefont {Iwata}},\ }\href
  {\doibase 10.1016/0022-0248(81)90206-2} {\bibfield  {journal} {\bibinfo
  {journal} {Journal of Crystal Growth}\ }\textbf {\bibinfo {volume} {52}},\
  \bibinfo {pages} {285} (\bibinfo {year} {1981})}\BibitemShut {NoStop}%
\bibitem [{\citenamefont {Hirano}\ \emph {et~al.}(1989)\citenamefont {Hirano},
  \citenamefont {Yogo}, \citenamefont {Asada},\ and\ \citenamefont
  {Naka}}]{hira1989}%
  \BibitemOpen
  \bibfield  {author} {\bibinfo {author} {\bibfnamefont {S.-I.}\ \bibnamefont
  {Hirano}}, \bibinfo {author} {\bibfnamefont {T.}~\bibnamefont {Yogo}},
  \bibinfo {author} {\bibfnamefont {S.}~\bibnamefont {Asada}}, \ and\ \bibinfo
  {author} {\bibfnamefont {S.}~\bibnamefont {Naka}},\ }\href {\doibase
  10.1111/j.1151-2916.1989.tb05955.x} {\bibfield  {journal} {\bibinfo
  {journal} {Journal of the American Ceramic Society}\ }\textbf {\bibinfo
  {volume} {72}},\ \bibinfo {pages} {66} (\bibinfo {year} {1989})}\BibitemShut
  {NoStop}%
\bibitem [{\citenamefont {Hamilton}\ \emph {et~al.}(1993)\citenamefont
  {Hamilton}, \citenamefont {Dolan}, \citenamefont {Mann}, \citenamefont
  {Colijn}, \citenamefont {McDonald},\ and\ \citenamefont {Shore}}]{ham1993}%
  \BibitemOpen
  \bibfield  {author} {\bibinfo {author} {\bibfnamefont {E.~J.~M.}\
  \bibnamefont {Hamilton}}, \bibinfo {author} {\bibfnamefont {S.~E.}\
  \bibnamefont {Dolan}}, \bibinfo {author} {\bibfnamefont {C.~M.}\ \bibnamefont
  {Mann}}, \bibinfo {author} {\bibfnamefont {H.~O.}\ \bibnamefont {Colijn}},
  \bibinfo {author} {\bibfnamefont {C.~A.}\ \bibnamefont {McDonald}}, \ and\
  \bibinfo {author} {\bibfnamefont {S.~G.}\ \bibnamefont {Shore}},\ }\href
  {\doibase 10.1126/science.260.5108.659} {\bibfield  {journal} {\bibinfo
  {journal} {Science}\ }\textbf {\bibinfo {volume} {260}},\ \bibinfo {pages}
  {659} (\bibinfo {year} {1993})}\BibitemShut {NoStop}%
\bibitem [{\citenamefont {Zedlitz}\ \emph {et~al.}(1996)\citenamefont
  {Zedlitz}, \citenamefont {Heintze},\ and\ \citenamefont
  {Schubert}}]{zed1996}%
  \BibitemOpen
  \bibfield  {author} {\bibinfo {author} {\bibfnamefont {R.}~\bibnamefont
  {Zedlitz}}, \bibinfo {author} {\bibfnamefont {M.}~\bibnamefont {Heintze}}, \
  and\ \bibinfo {author} {\bibfnamefont {M.}~\bibnamefont {Schubert}},\ }\href
  {\doibase 10.1016/0022-3093(95)00748-2} {\bibfield  {journal} {\bibinfo
  {journal} {Journal of Non-Crystalline Solids}\ }\textbf {\bibinfo {volume}
  {198-200}},\ \bibinfo {pages} {403} (\bibinfo {year} {1996})}\BibitemShut
  {NoStop}%
\bibitem [{\citenamefont {Balmain}(1842)}]{bal1842}%
  \BibitemOpen
  \bibfield  {author} {\bibinfo {author} {\bibfnamefont {W.~H.}\ \bibnamefont
  {Balmain}},\ }\href {\doibase 10.1002/prac.18420270164} {\bibfield  {journal}
  {\bibinfo  {journal} {Journal f{\"{u}}r Praktische Chemie}\ }\textbf
  {\bibinfo {volume} {27}},\ \bibinfo {pages} {422} (\bibinfo {year}
  {1842})}\BibitemShut {NoStop}%
\bibitem [{\citenamefont {Taylor}(1952)}]{tay1952}%
  \BibitemOpen
  \bibfield  {author} {\bibinfo {author} {\bibfnamefont {K.~M.}\ \bibnamefont
  {Taylor}},\ }\href@noop {} {} (\bibinfo {year} {1952})\BibitemShut {NoStop}%
\bibitem [{\citenamefont {Davis}(1991)}]{dav1991}%
  \BibitemOpen
  \bibfield  {author} {\bibinfo {author} {\bibfnamefont {R.}~\bibnamefont
  {Davis}},\ }\href {\doibase 10.1109/5.90133} {\bibfield  {journal} {\bibinfo
  {journal} {Proceedings of the IEEE}\ }\textbf {\bibinfo {volume} {79}},\
  \bibinfo {pages} {702} (\bibinfo {year} {1991})}\BibitemShut {NoStop}%
\bibitem [{\citenamefont {Haubner}\ \emph {et~al.}(2002)\citenamefont
  {Haubner}, \citenamefont {Wilhelm}, \citenamefont {Weissenbacher},\ and\
  \citenamefont {Lux}}]{hau2002}%
  \BibitemOpen
  \bibfield  {author} {\bibinfo {author} {\bibfnamefont {R.}~\bibnamefont
  {Haubner}}, \bibinfo {author} {\bibfnamefont {M.}~\bibnamefont {Wilhelm}},
  \bibinfo {author} {\bibfnamefont {R.}~\bibnamefont {Weissenbacher}}, \ and\
  \bibinfo {author} {\bibfnamefont {B.}~\bibnamefont {Lux}},\ }in\ \href
  {\doibase 10.1007/3-540-45623-6_1} {\emph {\bibinfo {booktitle} {High
  Performance Non-Oxide Ceramics II, Structure and Bonding}}},\ Vol.\ \bibinfo
  {volume} {102},\ \bibinfo {editor} {edited by\ \bibinfo {editor}
  {\bibfnamefont {M.}~\bibnamefont {Jansen}}}\ (\bibinfo  {publisher} {Springer
  Berlin Heidelberg},\ \bibinfo {year} {2002})\ pp.\ \bibinfo {pages} {1--45},\
  \Eprint {http://arxiv.org/abs/arXiv:1011.1669v3} {arXiv:1011.1669v3}
  \BibitemShut {NoStop}%
\bibitem [{\citenamefont {Lee}\ \emph {et~al.}(2006)\citenamefont {Lee},
  \citenamefont {Nakamura}, \citenamefont {Kume},\ and\ \citenamefont
  {Watari}}]{lee2006}%
  \BibitemOpen
  \bibfield  {author} {\bibinfo {author} {\bibfnamefont {S.~K.}\ \bibnamefont
  {Lee}}, \bibinfo {author} {\bibfnamefont {K.}~\bibnamefont {Nakamura}},
  \bibinfo {author} {\bibfnamefont {S.}~\bibnamefont {Kume}}, \ and\ \bibinfo
  {author} {\bibfnamefont {K.}~\bibnamefont {Watari}},\ }\href {\doibase
  10.4028/www.scientific.net/MSF.510-511.398} {\bibfield  {journal} {\bibinfo
  {journal} {Materials Science Forum}\ }\textbf {\bibinfo {volume} {510-511}},\
  \bibinfo {pages} {398} (\bibinfo {year} {2006})}\BibitemShut {NoStop}%
\bibitem [{\citenamefont {Eichler}\ and\ \citenamefont
  {Lesniak}(2008)}]{eic2008}%
  \BibitemOpen
  \bibfield  {author} {\bibinfo {author} {\bibfnamefont {J.}~\bibnamefont
  {Eichler}}\ and\ \bibinfo {author} {\bibfnamefont {C.}~\bibnamefont
  {Lesniak}},\ }\href {\doibase 10.1016/j.jeurceramsoc.2007.09.005} {\bibfield
  {journal} {\bibinfo  {journal} {Journal of the European Ceramic Society}\
  }\textbf {\bibinfo {volume} {28}},\ \bibinfo {pages} {1105} (\bibinfo {year}
  {2008})}\BibitemShut {NoStop}%
\bibitem [{\citenamefont {Withers}\ \emph {et~al.}(2015)\citenamefont
  {Withers}, \citenamefont {Bointon}, \citenamefont {Hudson}, \citenamefont
  {Craciun},\ and\ \citenamefont {Russo}}]{wit2015}%
  \BibitemOpen
  \bibfield  {author} {\bibinfo {author} {\bibfnamefont {F.}~\bibnamefont
  {Withers}}, \bibinfo {author} {\bibfnamefont {T.~H.}\ \bibnamefont
  {Bointon}}, \bibinfo {author} {\bibfnamefont {D.~C.}\ \bibnamefont {Hudson}},
  \bibinfo {author} {\bibfnamefont {M.~F.}\ \bibnamefont {Craciun}}, \ and\
  \bibinfo {author} {\bibfnamefont {S.}~\bibnamefont {Russo}},\ }\href
  {\doibase 10.1038/srep04967} {\bibfield  {journal} {\bibinfo  {journal}
  {Scientific Reports}\ }\textbf {\bibinfo {volume} {4}},\ \bibinfo {pages}
  {4967} (\bibinfo {year} {2015})}\BibitemShut {NoStop}%
\bibitem [{\citenamefont {Laturia}\ \emph {et~al.}(2018)\citenamefont
  {Laturia}, \citenamefont {{Van de Put}},\ and\ \citenamefont
  {Vandenberghe}}]{lat2018}%
  \BibitemOpen
  \bibfield  {author} {\bibinfo {author} {\bibfnamefont {A.}~\bibnamefont
  {Laturia}}, \bibinfo {author} {\bibfnamefont {M.~L.}\ \bibnamefont {{Van de
  Put}}}, \ and\ \bibinfo {author} {\bibfnamefont {W.~G.}\ \bibnamefont
  {Vandenberghe}},\ }\href {\doibase 10.1038/s41699-018-0050-x} {\bibfield
  {journal} {\bibinfo  {journal} {npj 2D Materials and Applications}\ }\textbf
  {\bibinfo {volume} {2}},\ \bibinfo {pages} {6} (\bibinfo {year}
  {2018})}\BibitemShut {NoStop}%
\bibitem [{\citenamefont {Bhagavantam}\ and\ \citenamefont {{Narayana
  Rao}}(1948)}]{bhag1948}%
  \BibitemOpen
  \bibfield  {author} {\bibinfo {author} {\bibfnamefont {S.}~\bibnamefont
  {Bhagavantam}}\ and\ \bibinfo {author} {\bibfnamefont {D.~A. A.~S.}\
  \bibnamefont {{Narayana Rao}}},\ }\href {\doibase 10.1038/161729a0}
  {\bibfield  {journal} {\bibinfo  {journal} {Nature}\ }\textbf {\bibinfo
  {volume} {161}},\ \bibinfo {pages} {729} (\bibinfo {year}
  {1948})}\BibitemShut {NoStop}%
\bibitem [{\citenamefont {Williams}(2011)}]{olie2011}%
  \BibitemOpen
  \bibfield  {author} {\bibinfo {author} {\bibfnamefont {O.}~\bibnamefont
  {Williams}},\ }\href {\doibase 10.1016/j.diamond.2011.02.015} {\bibfield
  {journal} {\bibinfo  {journal} {Diamond and Related Materials}\ }\textbf
  {\bibinfo {volume} {20}},\ \bibinfo {pages} {621} (\bibinfo {year}
  {2011})}\BibitemShut {NoStop}%
\bibitem [{\citenamefont {Harkins}(1942)}]{har1942}%
  \BibitemOpen
  \bibfield  {author} {\bibinfo {author} {\bibfnamefont {W.~D.}\ \bibnamefont
  {Harkins}},\ }\href {\doibase 10.1063/1.1723719} {\bibfield  {journal}
  {\bibinfo  {journal} {The Journal of Chemical Physics}\ }\textbf {\bibinfo
  {volume} {10}},\ \bibinfo {pages} {268} (\bibinfo {year} {1942})}\BibitemShut
  {NoStop}%
\bibitem [{\citenamefont {Seth}\ \emph {et~al.}(2002)\citenamefont {Seth},
  \citenamefont {Hatzikiriakos},\ and\ \citenamefont {Clere}}]{set2002}%
  \BibitemOpen
  \bibfield  {author} {\bibinfo {author} {\bibfnamefont {M.}~\bibnamefont
  {Seth}}, \bibinfo {author} {\bibfnamefont {S.~G.}\ \bibnamefont
  {Hatzikiriakos}}, \ and\ \bibinfo {author} {\bibfnamefont {T.~M.}\
  \bibnamefont {Clere}},\ }\href {\doibase 10.1002/pen.10986} {\bibfield
  {journal} {\bibinfo  {journal} {Polymer Engineering {\&} Science}\ }\textbf
  {\bibinfo {volume} {42}},\ \bibinfo {pages} {743} (\bibinfo {year}
  {2002})}\BibitemShut {NoStop}%
\bibitem [{\citenamefont {Hees}\ \emph {et~al.}(2011)\citenamefont {Hees},
  \citenamefont {Kriele},\ and\ \citenamefont {Williams}}]{hees2011}%
  \BibitemOpen
  \bibfield  {author} {\bibinfo {author} {\bibfnamefont {J.}~\bibnamefont
  {Hees}}, \bibinfo {author} {\bibfnamefont {A.}~\bibnamefont {Kriele}}, \ and\
  \bibinfo {author} {\bibfnamefont {O.~A.}\ \bibnamefont {Williams}},\ }\href
  {\doibase 10.1016/j.cplett.2011.04.083} {\bibfield  {journal} {\bibinfo
  {journal} {Chemical Physics Letters}\ }\textbf {\bibinfo {volume} {509}},\
  \bibinfo {pages} {12} (\bibinfo {year} {2011})}\BibitemShut {NoStop}%
\bibitem [{\citenamefont {{Van Wagenen}}\ and\ \citenamefont
  {Andrade}(1980)}]{wag1980}%
  \BibitemOpen
  \bibfield  {author} {\bibinfo {author} {\bibfnamefont {R.}~\bibnamefont {{Van
  Wagenen}}}\ and\ \bibinfo {author} {\bibfnamefont {J.}~\bibnamefont
  {Andrade}},\ }\href {\doibase 10.1016/0021-9797(80)90374-4} {\bibfield
  {journal} {\bibinfo  {journal} {Journal of Colloid and Interface Science}\
  }\textbf {\bibinfo {volume} {76}},\ \bibinfo {pages} {305} (\bibinfo {year}
  {1980})}\BibitemShut {NoStop}%
\bibitem [{\citenamefont {Voigt}\ \emph {et~al.}(1983)\citenamefont {Voigt},
  \citenamefont {Wolf}, \citenamefont {Lauekner}, \citenamefont {Neumann},
  \citenamefont {Becker},\ and\ \citenamefont {Richter}}]{voi1983}%
  \BibitemOpen
  \bibfield  {author} {\bibinfo {author} {\bibfnamefont {A.}~\bibnamefont
  {Voigt}}, \bibinfo {author} {\bibfnamefont {H.}~\bibnamefont {Wolf}},
  \bibinfo {author} {\bibfnamefont {S.}~\bibnamefont {Lauekner}}, \bibinfo
  {author} {\bibfnamefont {G.}~\bibnamefont {Neumann}}, \bibinfo {author}
  {\bibfnamefont {R.}~\bibnamefont {Becker}}, \ and\ \bibinfo {author}
  {\bibfnamefont {L.}~\bibnamefont {Richter}},\ }\href {\doibase
  10.1016/0142-9612(83)90032-7} {\bibfield  {journal} {\bibinfo  {journal}
  {Biomaterials}\ }\textbf {\bibinfo {volume} {4}},\ \bibinfo {pages} {299}
  (\bibinfo {year} {1983})}\BibitemShut {NoStop}%
\bibitem [{\citenamefont {Norde}\ and\ \citenamefont
  {Rouwendal}(1990)}]{nor1990}%
  \BibitemOpen
  \bibfield  {author} {\bibinfo {author} {\bibfnamefont {W.}~\bibnamefont
  {Norde}}\ and\ \bibinfo {author} {\bibfnamefont {E.}~\bibnamefont
  {Rouwendal}},\ }\href {\doibase 10.1016/0021-9797(90)90454-V} {\bibfield
  {journal} {\bibinfo  {journal} {Journal of Colloid and Interface Science}\
  }\textbf {\bibinfo {volume} {139}},\ \bibinfo {pages} {169} (\bibinfo {year}
  {1990})}\BibitemShut {NoStop}%
\bibitem [{\citenamefont {Scales}\ \emph {et~al.}(1990)\citenamefont {Scales},
  \citenamefont {Grieser},\ and\ \citenamefont {Healy}}]{sca1990}%
  \BibitemOpen
  \bibfield  {author} {\bibinfo {author} {\bibfnamefont {P.~J.}\ \bibnamefont
  {Scales}}, \bibinfo {author} {\bibfnamefont {F.}~\bibnamefont {Grieser}}, \
  and\ \bibinfo {author} {\bibfnamefont {T.~W.}\ \bibnamefont {Healy}},\ }\href
  {\doibase 10.1021/la00093a012} {\bibfield  {journal} {\bibinfo  {journal}
  {Langmuir}\ }\textbf {\bibinfo {volume} {6}},\ \bibinfo {pages} {582}
  (\bibinfo {year} {1990})}\BibitemShut {NoStop}%
\bibitem [{\citenamefont {Mandal}\ \emph {et~al.}(2017)\citenamefont {Mandal},
  \citenamefont {Thomas}, \citenamefont {Middleton}, \citenamefont {Gines},
  \citenamefont {Griffiths}, \citenamefont {Kappers}, \citenamefont {Oliver},
  \citenamefont {Wallis}, \citenamefont {Goff}, \citenamefont {Lynch},
  \citenamefont {Kuball},\ and\ \citenamefont {Williams}}]{man2017}%
  \BibitemOpen
  \bibfield  {author} {\bibinfo {author} {\bibfnamefont {S.}~\bibnamefont
  {Mandal}}, \bibinfo {author} {\bibfnamefont {E.~L.~H.}\ \bibnamefont
  {Thomas}}, \bibinfo {author} {\bibfnamefont {C.}~\bibnamefont {Middleton}},
  \bibinfo {author} {\bibfnamefont {L.}~\bibnamefont {Gines}}, \bibinfo
  {author} {\bibfnamefont {J.~T.}\ \bibnamefont {Griffiths}}, \bibinfo {author}
  {\bibfnamefont {M.~J.}\ \bibnamefont {Kappers}}, \bibinfo {author}
  {\bibfnamefont {R.~A.}\ \bibnamefont {Oliver}}, \bibinfo {author}
  {\bibfnamefont {D.~J.}\ \bibnamefont {Wallis}}, \bibinfo {author}
  {\bibfnamefont {L.~E.}\ \bibnamefont {Goff}}, \bibinfo {author}
  {\bibfnamefont {S.~A.}\ \bibnamefont {Lynch}}, \bibinfo {author}
  {\bibfnamefont {M.}~\bibnamefont {Kuball}}, \ and\ \bibinfo {author}
  {\bibfnamefont {O.~A.}\ \bibnamefont {Williams}},\ }\href {\doibase
  10.1021/acsomega.7b01069} {\bibfield  {journal} {\bibinfo  {journal} {ACS
  Omega}\ }\textbf {\bibinfo {volume} {2}},\ \bibinfo {pages} {7275} (\bibinfo
  {year} {2017})},\ \Eprint {http://arxiv.org/abs/1707.05410} {1707.05410}
  \BibitemShut {NoStop}%
\bibitem [{\citenamefont {Klemencic}\ \emph
  {et~al.}(2017{\natexlab{a}})\citenamefont {Klemencic}, \citenamefont
  {Mandal}, \citenamefont {Werrell}, \citenamefont {Giblin},\ and\
  \citenamefont {Williams}}]{klemp2017}%
  \BibitemOpen
  \bibfield  {author} {\bibinfo {author} {\bibfnamefont {G.~M.}\ \bibnamefont
  {Klemencic}}, \bibinfo {author} {\bibfnamefont {S.}~\bibnamefont {Mandal}},
  \bibinfo {author} {\bibfnamefont {J.~M.}\ \bibnamefont {Werrell}}, \bibinfo
  {author} {\bibfnamefont {S.~R.}\ \bibnamefont {Giblin}}, \ and\ \bibinfo
  {author} {\bibfnamefont {O.~A.}\ \bibnamefont {Williams}},\ }\href {\doibase
  10.1080/14686996.2017.1286223} {\bibfield  {journal} {\bibinfo  {journal}
  {Science and Technology of Advanced Materials}\ }\textbf {\bibinfo {volume}
  {18}},\ \bibinfo {pages} {239} (\bibinfo {year}
  {2017}{\natexlab{a}})}\BibitemShut {NoStop}%
\bibitem [{\citenamefont {Klemencic}\ \emph
  {et~al.}(2017{\natexlab{b}})\citenamefont {Klemencic}, \citenamefont
  {Fellows}, \citenamefont {Werrell}, \citenamefont {Mandal}, \citenamefont
  {Giblin}, \citenamefont {Smith},\ and\ \citenamefont {Williams}}]{klem2017}%
  \BibitemOpen
  \bibfield  {author} {\bibinfo {author} {\bibfnamefont {G.~M.}\ \bibnamefont
  {Klemencic}}, \bibinfo {author} {\bibfnamefont {J.~M.}\ \bibnamefont
  {Fellows}}, \bibinfo {author} {\bibfnamefont {J.~M.}\ \bibnamefont
  {Werrell}}, \bibinfo {author} {\bibfnamefont {S.}~\bibnamefont {Mandal}},
  \bibinfo {author} {\bibfnamefont {S.~R.}\ \bibnamefont {Giblin}}, \bibinfo
  {author} {\bibfnamefont {R.~A.}\ \bibnamefont {Smith}}, \ and\ \bibinfo
  {author} {\bibfnamefont {O.~A.}\ \bibnamefont {Williams}},\ }\href {\doibase
  10.1103/PhysRevMaterials.1.044801} {\bibfield  {journal} {\bibinfo  {journal}
  {Physical Review Materials}\ }\textbf {\bibinfo {volume} {1}},\ \bibinfo
  {pages} {044801} (\bibinfo {year} {2017}{\natexlab{b}})}\BibitemShut
  {NoStop}%
\bibitem [{\citenamefont {Bland}\ \emph {et~al.}(2018)\citenamefont {Bland},
  \citenamefont {Thomas}, \citenamefont {Klemencic}, \citenamefont {Mandal},
  \citenamefont {Papageorgiou}, \citenamefont {Jones},\ and\ \citenamefont
  {Williams}}]{bla2018}%
  \BibitemOpen
  \bibfield  {author} {\bibinfo {author} {\bibfnamefont {H.~A.}\ \bibnamefont
  {Bland}}, \bibinfo {author} {\bibfnamefont {E.~L.~H.}\ \bibnamefont
  {Thomas}}, \bibinfo {author} {\bibfnamefont {G.~M.}\ \bibnamefont
  {Klemencic}}, \bibinfo {author} {\bibfnamefont {S.}~\bibnamefont {Mandal}},
  \bibinfo {author} {\bibfnamefont {A.}~\bibnamefont {Papageorgiou}}, \bibinfo
  {author} {\bibfnamefont {T.~G.}\ \bibnamefont {Jones}}, \ and\ \bibinfo
  {author} {\bibfnamefont {O.~A.}\ \bibnamefont {Williams}},\ }\href {\doibase
  arXiv:1810.10282v1} {\bibfield  {journal} {\bibinfo  {journal} {ArXiv}\ ,\
  \bibinfo {pages} {1810.10282}} (\bibinfo {year} {2018})},\ \Eprint
  {http://arxiv.org/abs/1810.10282} {arXiv:1810.10282} \BibitemShut {NoStop}%
\bibitem [{\citenamefont {Leeds}\ \emph {et~al.}(1998)\citenamefont {Leeds},
  \citenamefont {Davis}, \citenamefont {May}, \citenamefont {Pickard},\ and\
  \citenamefont {Ashfold}}]{leeds1998}%
  \BibitemOpen
  \bibfield  {author} {\bibinfo {author} {\bibfnamefont {S.}~\bibnamefont
  {Leeds}}, \bibinfo {author} {\bibfnamefont {T.}~\bibnamefont {Davis}},
  \bibinfo {author} {\bibfnamefont {P.}~\bibnamefont {May}}, \bibinfo {author}
  {\bibfnamefont {C.}~\bibnamefont {Pickard}}, \ and\ \bibinfo {author}
  {\bibfnamefont {M.}~\bibnamefont {Ashfold}},\ }\href {\doibase
  10.1016/S0925-9635(97)00261-6} {\bibfield  {journal} {\bibinfo  {journal}
  {Diamond and Related Materials}\ }\textbf {\bibinfo {volume} {7}},\ \bibinfo
  {pages} {233} (\bibinfo {year} {1998})}\BibitemShut {NoStop}%
\bibitem [{\citenamefont {Cuenca}\ \emph {et~al.}(2019)\citenamefont {Cuenca},
  \citenamefont {Mandal}, \citenamefont {Porch},\ and\ \citenamefont
  {Williams}}]{cue2019}%
  \BibitemOpen
  \bibfield  {author} {\bibinfo {author} {\bibfnamefont {J.~A.}\ \bibnamefont
  {Cuenca}}, \bibinfo {author} {\bibfnamefont {S.}~\bibnamefont {Mandal}},
  \bibinfo {author} {\bibfnamefont {A.}~\bibnamefont {Porch}}, \ and\ \bibinfo
  {author} {\bibfnamefont {O.~A.}\ \bibnamefont {Williams}},\ }\href@noop {}
  {\bibfield  {journal} {\bibinfo  {journal} {Under Preparation}\ ,\ \bibinfo
  {pages} {xx}} (\bibinfo {year} {2019})}\BibitemShut {NoStop}%
\bibitem [{\citenamefont {Williams}\ \emph {et~al.}(2010)\citenamefont
  {Williams}, \citenamefont {Hees}, \citenamefont {Dieker}, \citenamefont
  {J{\"{a}}ger}, \citenamefont {Kirste},\ and\ \citenamefont
  {Nebel}}]{hees2010}%
  \BibitemOpen
  \bibfield  {author} {\bibinfo {author} {\bibfnamefont {O.~A.}\ \bibnamefont
  {Williams}}, \bibinfo {author} {\bibfnamefont {J.}~\bibnamefont {Hees}},
  \bibinfo {author} {\bibfnamefont {C.}~\bibnamefont {Dieker}}, \bibinfo
  {author} {\bibfnamefont {W.}~\bibnamefont {J{\"{a}}ger}}, \bibinfo {author}
  {\bibfnamefont {L.}~\bibnamefont {Kirste}}, \ and\ \bibinfo {author}
  {\bibfnamefont {C.~E.}\ \bibnamefont {Nebel}},\ }\href {\doibase
  10.1021/nn100748k} {\bibfield  {journal} {\bibinfo  {journal} {ACS Nano}\
  }\textbf {\bibinfo {volume} {4}},\ \bibinfo {pages} {4824} (\bibinfo {year}
  {2010})}\BibitemShut {NoStop}%
\bibitem [{\citenamefont {Crimp}\ \emph {et~al.}(1999)\citenamefont {Crimp},
  \citenamefont {Oppermann},\ and\ \citenamefont {Krehbiel}}]{crimp1999}%
  \BibitemOpen
  \bibfield  {author} {\bibinfo {author} {\bibfnamefont {M.~J.}\ \bibnamefont
  {Crimp}}, \bibinfo {author} {\bibfnamefont {D.~A.}\ \bibnamefont
  {Oppermann}}, \ and\ \bibinfo {author} {\bibfnamefont {K.}~\bibnamefont
  {Krehbiel}},\ }\href {\doibase 10.1023/A:1004656817379} {\bibfield  {journal}
  {\bibinfo  {journal} {Journal of Materials Science}\ }\textbf {\bibinfo
  {volume} {34}},\ \bibinfo {pages} {2621} (\bibinfo {year}
  {1999})}\BibitemShut {NoStop}%
\bibitem [{\citenamefont {Li}\ \emph {et~al.}(2015)\citenamefont {Li},
  \citenamefont {Huang}, \citenamefont {Liu}, \citenamefont {Zhang},
  \citenamefont {Liu}, \citenamefont {Luo}, \citenamefont {Ma}, \citenamefont
  {Xu}, \citenamefont {Lu}, \citenamefont {Lin}, \citenamefont {Zou},\ and\
  \citenamefont {Tang}}]{li2015}%
  \BibitemOpen
  \bibfield  {author} {\bibinfo {author} {\bibfnamefont {J.}~\bibnamefont
  {Li}}, \bibinfo {author} {\bibfnamefont {Y.}~\bibnamefont {Huang}}, \bibinfo
  {author} {\bibfnamefont {Z.}~\bibnamefont {Liu}}, \bibinfo {author}
  {\bibfnamefont {J.}~\bibnamefont {Zhang}}, \bibinfo {author} {\bibfnamefont
  {X.}~\bibnamefont {Liu}}, \bibinfo {author} {\bibfnamefont {H.}~\bibnamefont
  {Luo}}, \bibinfo {author} {\bibfnamefont {Y.}~\bibnamefont {Ma}}, \bibinfo
  {author} {\bibfnamefont {X.}~\bibnamefont {Xu}}, \bibinfo {author}
  {\bibfnamefont {Y.}~\bibnamefont {Lu}}, \bibinfo {author} {\bibfnamefont
  {J.}~\bibnamefont {Lin}}, \bibinfo {author} {\bibfnamefont {J.}~\bibnamefont
  {Zou}}, \ and\ \bibinfo {author} {\bibfnamefont {C.}~\bibnamefont {Tang}},\
  }\href {\doibase 10.1039/C5TA00601E} {\bibfield  {journal} {\bibinfo
  {journal} {Journal of Materials Chemistry A}\ }\textbf {\bibinfo {volume}
  {3}},\ \bibinfo {pages} {8185} (\bibinfo {year} {2015})}\BibitemShut
  {NoStop}%
\bibitem [{\citenamefont {Williams}\ and\ \citenamefont
  {Hawn}(1991)}]{dou1991}%
  \BibitemOpen
  \bibfield  {author} {\bibinfo {author} {\bibfnamefont {P.~D.}\ \bibnamefont
  {Williams}}\ and\ \bibinfo {author} {\bibfnamefont {D.~D.}\ \bibnamefont
  {Hawn}},\ }\href {\doibase 10.1111/j.1151-2916.1991.tb07147.x} {\bibfield
  {journal} {\bibinfo  {journal} {Journal of the American Ceramic Society}\
  }\textbf {\bibinfo {volume} {74}},\ \bibinfo {pages} {1614} (\bibinfo {year}
  {1991})}\BibitemShut {NoStop}%
\bibitem [{\citenamefont {Lewis}(2004)}]{lewis2004}%
  \BibitemOpen
  \bibfield  {author} {\bibinfo {author} {\bibfnamefont {J.~A.}\ \bibnamefont
  {Lewis}},\ }\href {\doibase 10.1111/j.1151-2916.2000.tb01560.x} {\bibfield
  {journal} {\bibinfo  {journal} {Journal of the American Ceramic Society}\
  }\textbf {\bibinfo {volume} {83}},\ \bibinfo {pages} {2341} (\bibinfo {year}
  {2004})},\ \Eprint {http://arxiv.org/abs/arXiv:1011.1669v3}
  {arXiv:arXiv:1011.1669v3} \BibitemShut {NoStop}%
\bibitem [{\citenamefont {Reich}\ \emph {et~al.}(2005)\citenamefont {Reich},
  \citenamefont {Ferrari}, \citenamefont {Arenal}, \citenamefont {Loiseau},
  \citenamefont {Bello},\ and\ \citenamefont {Robertson}}]{reich2005}%
  \BibitemOpen
  \bibfield  {author} {\bibinfo {author} {\bibfnamefont {S.}~\bibnamefont
  {Reich}}, \bibinfo {author} {\bibfnamefont {A.~C.}\ \bibnamefont {Ferrari}},
  \bibinfo {author} {\bibfnamefont {R.}~\bibnamefont {Arenal}}, \bibinfo
  {author} {\bibfnamefont {A.}~\bibnamefont {Loiseau}}, \bibinfo {author}
  {\bibfnamefont {I.}~\bibnamefont {Bello}}, \ and\ \bibinfo {author}
  {\bibfnamefont {J.}~\bibnamefont {Robertson}},\ }\href {\doibase
  10.1103/PhysRevB.71.205201} {\bibfield  {journal} {\bibinfo  {journal}
  {Physical Review B}\ }\textbf {\bibinfo {volume} {71}},\ \bibinfo {pages}
  {205201} (\bibinfo {year} {2005})}\BibitemShut {NoStop}%
\bibitem [{\citenamefont {Ramaswamy}(1930)}]{ram1930}%
  \BibitemOpen
  \bibfield  {author} {\bibinfo {author} {\bibfnamefont {C.}~\bibnamefont
  {Ramaswamy}},\ }\href {\doibase 10.1038/125704b0} {\bibfield  {journal}
  {\bibinfo  {journal} {Nature}\ }\textbf {\bibinfo {volume} {125}},\ \bibinfo
  {pages} {704} (\bibinfo {year} {1930})}\BibitemShut {NoStop}%
\bibitem [{\citenamefont {Robertson}\ and\ \citenamefont
  {Fox}(1930)}]{robert1930}%
  \BibitemOpen
  \bibfield  {author} {\bibinfo {author} {\bibfnamefont {R.}~\bibnamefont
  {Robertson}}\ and\ \bibinfo {author} {\bibfnamefont {J.~J.}\ \bibnamefont
  {Fox}},\ }\href {\doibase 10.1038/125704a0} {\bibfield  {journal} {\bibinfo
  {journal} {Nature}\ }\textbf {\bibinfo {volume} {125}},\ \bibinfo {pages}
  {704} (\bibinfo {year} {1930})}\BibitemShut {NoStop}%
\bibitem [{\citenamefont {Bhagavantam}(1930)}]{bhag1930}%
  \BibitemOpen
  \bibfield  {author} {\bibinfo {author} {\bibfnamefont {S.}~\bibnamefont
  {Bhagavantam}},\ }\href {http://hdl.handle.net/10821/555} {\bibfield
  {journal} {\bibinfo  {journal} {Indian Journal of Physics}\ }\textbf
  {\bibinfo {volume} {5}},\ \bibinfo {pages} {169} (\bibinfo {year}
  {1930})}\BibitemShut {NoStop}%
\bibitem [{\citenamefont {May}\ \emph {et~al.}(2008)\citenamefont {May},
  \citenamefont {Ludlow}, \citenamefont {Hannaway}, \citenamefont {Heard},
  \citenamefont {Smith},\ and\ \citenamefont {Rosser}}]{may2008}%
  \BibitemOpen
  \bibfield  {author} {\bibinfo {author} {\bibfnamefont {P.}~\bibnamefont
  {May}}, \bibinfo {author} {\bibfnamefont {W.}~\bibnamefont {Ludlow}},
  \bibinfo {author} {\bibfnamefont {M.}~\bibnamefont {Hannaway}}, \bibinfo
  {author} {\bibfnamefont {P.}~\bibnamefont {Heard}}, \bibinfo {author}
  {\bibfnamefont {J.}~\bibnamefont {Smith}}, \ and\ \bibinfo {author}
  {\bibfnamefont {K.}~\bibnamefont {Rosser}},\ }\href {\doibase
  10.1016/j.diamond.2007.11.005} {\bibfield  {journal} {\bibinfo  {journal}
  {Diamond and Related Materials}\ }\textbf {\bibinfo {volume} {17}},\ \bibinfo
  {pages} {105} (\bibinfo {year} {2008})}\BibitemShut {NoStop}%
\bibitem [{\citenamefont {Szirmai}\ \emph {et~al.}(2012)\citenamefont
  {Szirmai}, \citenamefont {Pichler}, \citenamefont {Williams}, \citenamefont
  {Mandal}, \citenamefont {B{\"{a}}uerle},\ and\ \citenamefont
  {Simon}}]{szi2012}%
  \BibitemOpen
  \bibfield  {author} {\bibinfo {author} {\bibfnamefont {P.}~\bibnamefont
  {Szirmai}}, \bibinfo {author} {\bibfnamefont {T.}~\bibnamefont {Pichler}},
  \bibinfo {author} {\bibfnamefont {O.~a.}\ \bibnamefont {Williams}}, \bibinfo
  {author} {\bibfnamefont {S.}~\bibnamefont {Mandal}}, \bibinfo {author}
  {\bibfnamefont {C.}~\bibnamefont {B{\"{a}}uerle}}, \ and\ \bibinfo {author}
  {\bibfnamefont {F.}~\bibnamefont {Simon}},\ }\href {\doibase
  10.1002/pssb.201200461} {\bibfield  {journal} {\bibinfo  {journal} {Physica
  Status Solidi (b)}\ }\textbf {\bibinfo {volume} {249}},\ \bibinfo {pages}
  {2656} (\bibinfo {year} {2012})}\BibitemShut {NoStop}%
\bibitem [{\citenamefont {Sidorov}\ and\ \citenamefont
  {Ekimov}(2010)}]{sidorov2010}%
  \BibitemOpen
  \bibfield  {author} {\bibinfo {author} {\bibfnamefont {V.~A.}\ \bibnamefont
  {Sidorov}}\ and\ \bibinfo {author} {\bibfnamefont {E.~A.}\ \bibnamefont
  {Ekimov}},\ }\href {\doibase 10.1016/j.diamond.2009.12.002} {\bibfield
  {journal} {\bibinfo  {journal} {Diamond and Related Materials}\ }\textbf
  {\bibinfo {volume} {19}},\ \bibinfo {pages} {351} (\bibinfo {year} {2010})},\
  \Eprint {http://arxiv.org/abs/0404156} {arXiv:0404156 [cond-mat]}
  \BibitemShut {NoStop}%
\bibitem [{\citenamefont {Ashcheulov}\ \emph {et~al.}(2013)\citenamefont
  {Ashcheulov}, \citenamefont {{\v{S}}ebera}, \citenamefont {Kovalenko},
  \citenamefont {Petr{\'{a}}k}, \citenamefont {Fendrych}, \citenamefont
  {Nesl{\'{a}}dek}, \citenamefont {Taylor}, \citenamefont
  {{\v{Z}}ivcov{\'{a}}}, \citenamefont {Frank}, \citenamefont {Kavan},
  \citenamefont {Dra{\v{c}}{\'{i}}nsk{\'{y}}}, \citenamefont {Hub{\'{i}}k},
  \citenamefont {Vac{\'{i}}k}, \citenamefont {Kraus},\ and\ \citenamefont
  {Kratochv{\'{i}}lov{\'{a}}}}]{ash2013}%
  \BibitemOpen
  \bibfield  {author} {\bibinfo {author} {\bibfnamefont {P.}~\bibnamefont
  {Ashcheulov}}, \bibinfo {author} {\bibfnamefont {J.}~\bibnamefont
  {{\v{S}}ebera}}, \bibinfo {author} {\bibfnamefont {A.}~\bibnamefont
  {Kovalenko}}, \bibinfo {author} {\bibfnamefont {V.}~\bibnamefont
  {Petr{\'{a}}k}}, \bibinfo {author} {\bibfnamefont {F.}~\bibnamefont
  {Fendrych}}, \bibinfo {author} {\bibfnamefont {M.}~\bibnamefont
  {Nesl{\'{a}}dek}}, \bibinfo {author} {\bibfnamefont {A.}~\bibnamefont
  {Taylor}}, \bibinfo {author} {\bibfnamefont {Z.~V.}\ \bibnamefont
  {{\v{Z}}ivcov{\'{a}}}}, \bibinfo {author} {\bibfnamefont {O.}~\bibnamefont
  {Frank}}, \bibinfo {author} {\bibfnamefont {L.}~\bibnamefont {Kavan}},
  \bibinfo {author} {\bibfnamefont {M.}~\bibnamefont
  {Dra{\v{c}}{\'{i}}nsk{\'{y}}}}, \bibinfo {author} {\bibfnamefont
  {P.}~\bibnamefont {Hub{\'{i}}k}}, \bibinfo {author} {\bibfnamefont
  {J.}~\bibnamefont {Vac{\'{i}}k}}, \bibinfo {author} {\bibfnamefont
  {I.}~\bibnamefont {Kraus}}, \ and\ \bibinfo {author} {\bibfnamefont
  {I.}~\bibnamefont {Kratochv{\'{i}}lov{\'{a}}}},\ }\href {\doibase
  10.1140/epjb/e2013-40528-x} {\bibfield  {journal} {\bibinfo  {journal}
  {European Physical Journal B}\ }\textbf {\bibinfo {volume} {86}},\ \bibinfo
  {pages} {12} (\bibinfo {year} {2013})}\BibitemShut {NoStop}%
\bibitem [{\citenamefont {May}\ \emph {et~al.}(2007)\citenamefont {May},
  \citenamefont {Ludlow}, \citenamefont {Hannaway}, \citenamefont {Heard},
  \citenamefont {Smith},\ and\ \citenamefont {Rosser}}]{may2007}%
  \BibitemOpen
  \bibfield  {author} {\bibinfo {author} {\bibfnamefont {P.}~\bibnamefont
  {May}}, \bibinfo {author} {\bibfnamefont {W.}~\bibnamefont {Ludlow}},
  \bibinfo {author} {\bibfnamefont {M.}~\bibnamefont {Hannaway}}, \bibinfo
  {author} {\bibfnamefont {P.}~\bibnamefont {Heard}}, \bibinfo {author}
  {\bibfnamefont {J.}~\bibnamefont {Smith}}, \ and\ \bibinfo {author}
  {\bibfnamefont {K.}~\bibnamefont {Rosser}},\ }\href {\doibase
  10.1016/j.cplett.2007.08.018} {\bibfield  {journal} {\bibinfo  {journal}
  {Chemical Physics Letters}\ }\textbf {\bibinfo {volume} {446}},\ \bibinfo
  {pages} {103} (\bibinfo {year} {2007})}\BibitemShut {NoStop}%
\bibitem [{\citenamefont {Gheeraert}\ \emph {et~al.}(1993)\citenamefont
  {Gheeraert}, \citenamefont {Gonon}, \citenamefont {Deneuville}, \citenamefont
  {Abello},\ and\ \citenamefont {Lucazeau}}]{ghee1993}%
  \BibitemOpen
  \bibfield  {author} {\bibinfo {author} {\bibfnamefont {E.}~\bibnamefont
  {Gheeraert}}, \bibinfo {author} {\bibfnamefont {P.}~\bibnamefont {Gonon}},
  \bibinfo {author} {\bibfnamefont {A.}~\bibnamefont {Deneuville}}, \bibinfo
  {author} {\bibfnamefont {L.}~\bibnamefont {Abello}}, \ and\ \bibinfo {author}
  {\bibfnamefont {G.}~\bibnamefont {Lucazeau}},\ }\href {\doibase
  10.1016/0925-9635(93)90215-N} {\bibfield  {journal} {\bibinfo  {journal}
  {Diamond and Related Materials}\ }\textbf {\bibinfo {volume} {2}},\ \bibinfo
  {pages} {742} (\bibinfo {year} {1993})}\BibitemShut {NoStop}%
\bibitem [{\citenamefont {Ager}\ \emph {et~al.}(1995)\citenamefont {Ager},
  \citenamefont {Walukiewicz}, \citenamefont {McCluskey}, \citenamefont
  {Plano},\ and\ \citenamefont {Landstrass}}]{ager1995}%
  \BibitemOpen
  \bibfield  {author} {\bibinfo {author} {\bibfnamefont {J.~W.}\ \bibnamefont
  {Ager}}, \bibinfo {author} {\bibfnamefont {W.}~\bibnamefont {Walukiewicz}},
  \bibinfo {author} {\bibfnamefont {M.}~\bibnamefont {McCluskey}}, \bibinfo
  {author} {\bibfnamefont {M.~A.}\ \bibnamefont {Plano}}, \ and\ \bibinfo
  {author} {\bibfnamefont {M.~I.}\ \bibnamefont {Landstrass}},\ }\href
  {\doibase 10.1063/1.114031} {\bibfield  {journal} {\bibinfo  {journal}
  {Applied Physics Letters}\ }\textbf {\bibinfo {volume} {616}},\ \bibinfo
  {pages} {616} (\bibinfo {year} {1995})}\BibitemShut {NoStop}%
\bibitem [{\citenamefont {Pruvost}\ and\ \citenamefont
  {Deneuville}(2001)}]{pru2001}%
  \BibitemOpen
  \bibfield  {author} {\bibinfo {author} {\bibfnamefont {F.}~\bibnamefont
  {Pruvost}}\ and\ \bibinfo {author} {\bibfnamefont {A.}~\bibnamefont
  {Deneuville}},\ }\href {\doibase 10.1016/S0925-9635(00)00378-2} {\bibfield
  {journal} {\bibinfo  {journal} {Diamond and Related Materials}\ }\textbf
  {\bibinfo {volume} {10}},\ \bibinfo {pages} {531} (\bibinfo {year}
  {2001})}\BibitemShut {NoStop}%
\bibitem [{\citenamefont {Ferrari}\ and\ \citenamefont
  {Robertson}(2001{\natexlab{a}})}]{fer2001}%
  \BibitemOpen
  \bibfield  {author} {\bibinfo {author} {\bibfnamefont {A.~C.}\ \bibnamefont
  {Ferrari}}\ and\ \bibinfo {author} {\bibfnamefont {J.}~\bibnamefont
  {Robertson}},\ }\href {\doibase 10.1103/PhysRevB.63.121405} {\bibfield
  {journal} {\bibinfo  {journal} {Physical Review B - Condensed Matter and
  Materials Physics}\ }\textbf {\bibinfo {volume} {63}},\ \bibinfo {pages} {2}
  (\bibinfo {year} {2001}{\natexlab{a}})}\BibitemShut {NoStop}%
\bibitem [{\citenamefont {Ferrari}\ and\ \citenamefont
  {Robertson}(2000)}]{fer2000}%
  \BibitemOpen
  \bibfield  {author} {\bibinfo {author} {\bibfnamefont {A.~C.}\ \bibnamefont
  {Ferrari}}\ and\ \bibinfo {author} {\bibfnamefont {J.}~\bibnamefont
  {Robertson}},\ }\href {\doibase 10.1103/PhysRevB.61.14095} {\bibfield
  {journal} {\bibinfo  {journal} {Physical Review B}\ }\textbf {\bibinfo
  {volume} {61}},\ \bibinfo {pages} {14095} (\bibinfo {year}
  {2000})}\BibitemShut {NoStop}%
\bibitem [{\citenamefont {Ferrari}\ and\ \citenamefont
  {Robertson}(2001{\natexlab{b}})}]{fer2001a}%
  \BibitemOpen
  \bibfield  {author} {\bibinfo {author} {\bibfnamefont {A.~C.}\ \bibnamefont
  {Ferrari}}\ and\ \bibinfo {author} {\bibfnamefont {J.}~\bibnamefont
  {Robertson}},\ }\href {\doibase 10.1103/PhysRevB.64.075414} {\bibfield
  {journal} {\bibinfo  {journal} {Physical Review B}\ }\textbf {\bibinfo
  {volume} {64}},\ \bibinfo {pages} {075414} (\bibinfo {year}
  {2001}{\natexlab{b}})}\BibitemShut {NoStop}%
\bibitem [{\citenamefont {Ferrari}\ and\ \citenamefont
  {Robertson}(2004)}]{fer2004}%
  \BibitemOpen
  \bibfield  {author} {\bibinfo {author} {\bibfnamefont {A.~C.}\ \bibnamefont
  {Ferrari}}\ and\ \bibinfo {author} {\bibfnamefont {J.}~\bibnamefont
  {Robertson}},\ }\href {\doibase 10.1098/rsta.2004.1452} {\bibfield  {journal}
  {\bibinfo  {journal} {Philosophical Transactions of the Royal Society A:
  Mathematical, Physical and Engineering Sciences}\ }\textbf {\bibinfo {volume}
  {362}},\ \bibinfo {pages} {2477} (\bibinfo {year} {2004})}\BibitemShut
  {NoStop}%
\bibitem [{\citenamefont {Prawer}\ and\ \citenamefont
  {Nemanich}(2004)}]{pra2004}%
  \BibitemOpen
  \bibfield  {author} {\bibinfo {author} {\bibfnamefont {S.}~\bibnamefont
  {Prawer}}\ and\ \bibinfo {author} {\bibfnamefont {R.~J.}\ \bibnamefont
  {Nemanich}},\ }\href {\doibase 10.1098/rsta.2004.1451} {\bibfield  {journal}
  {\bibinfo  {journal} {Philosophical Transactions of the Royal Society of
  London. Series A: Mathematical, Physical and Engineering Sciences}\ }\textbf
  {\bibinfo {volume} {362}},\ \bibinfo {pages} {2537} (\bibinfo {year}
  {2004})}\BibitemShut {NoStop}%
\bibitem [{\citenamefont {Prawer}\ \emph {et~al.}(1998)\citenamefont {Prawer},
  \citenamefont {Nugent},\ and\ \citenamefont {Jamieson}}]{pra1998}%
  \BibitemOpen
  \bibfield  {author} {\bibinfo {author} {\bibfnamefont {S.}~\bibnamefont
  {Prawer}}, \bibinfo {author} {\bibfnamefont {K.}~\bibnamefont {Nugent}}, \
  and\ \bibinfo {author} {\bibfnamefont {D.}~\bibnamefont {Jamieson}},\ }\href
  {\doibase 10.1016/S0925-9635(97)00194-5} {\bibfield  {journal} {\bibinfo
  {journal} {Diamond and Related Materials}\ }\textbf {\bibinfo {volume} {7}},\
  \bibinfo {pages} {106} (\bibinfo {year} {1998})}\BibitemShut {NoStop}%
\bibitem [{\citenamefont {Xia}\ \emph {et~al.}(2017)\citenamefont {Xia},
  \citenamefont {Zhong},\ and\ \citenamefont {Weng}}]{xia2017}%
  \BibitemOpen
  \bibfield  {author} {\bibinfo {author} {\bibfnamefont {X.}~\bibnamefont
  {Xia}}, \bibinfo {author} {\bibfnamefont {Z.}~\bibnamefont {Zhong}}, \ and\
  \bibinfo {author} {\bibfnamefont {G.~J.}\ \bibnamefont {Weng}},\ }\href
  {\doibase 10.1016/j.mechmat.2017.03.014} {\bibfield  {journal} {\bibinfo
  {journal} {Mechanics of Materials}\ }\textbf {\bibinfo {volume} {109}},\
  \bibinfo {pages} {42} (\bibinfo {year} {2017})}\BibitemShut {NoStop}%
\bibitem [{\citenamefont {Cuenca}\ \emph {et~al.}(2018)\citenamefont {Cuenca},
  \citenamefont {Thomas}, \citenamefont {Mandal}, \citenamefont {Morgan},
  \citenamefont {Lloret}, \citenamefont {Araujo}, \citenamefont {Williams},\
  and\ \citenamefont {Porch}}]{cue2018}%
  \BibitemOpen
  \bibfield  {author} {\bibinfo {author} {\bibfnamefont {J.~A.}\ \bibnamefont
  {Cuenca}}, \bibinfo {author} {\bibfnamefont {E.~L.~H.}\ \bibnamefont
  {Thomas}}, \bibinfo {author} {\bibfnamefont {S.}~\bibnamefont {Mandal}},
  \bibinfo {author} {\bibfnamefont {D.~J.}\ \bibnamefont {Morgan}}, \bibinfo
  {author} {\bibfnamefont {F.}~\bibnamefont {Lloret}}, \bibinfo {author}
  {\bibfnamefont {D.}~\bibnamefont {Araujo}}, \bibinfo {author} {\bibfnamefont
  {O.~A.}\ \bibnamefont {Williams}}, \ and\ \bibinfo {author} {\bibfnamefont
  {A.}~\bibnamefont {Porch}},\ }\href {\doibase 10.1021/acsomega.7b02000}
  {\bibfield  {journal} {\bibinfo  {journal} {ACS Omega}\ }\textbf {\bibinfo
  {volume} {3}},\ \bibinfo {pages} {2183} (\bibinfo {year} {2018})}\BibitemShut
  {NoStop}%
\bibitem [{\citenamefont {Ekimov}\ \emph {et~al.}(2004)\citenamefont {Ekimov},
  \citenamefont {Sidorov}, \citenamefont {Bauer}, \citenamefont {Mel'nik},
  \citenamefont {Curro}, \citenamefont {Thompson},\ and\ \citenamefont
  {Stishov}}]{ekimov2004}%
  \BibitemOpen
  \bibfield  {author} {\bibinfo {author} {\bibfnamefont {E.~A.}\ \bibnamefont
  {Ekimov}}, \bibinfo {author} {\bibfnamefont {V.~A.}\ \bibnamefont {Sidorov}},
  \bibinfo {author} {\bibfnamefont {E.~D.}\ \bibnamefont {Bauer}}, \bibinfo
  {author} {\bibfnamefont {N.~N.}\ \bibnamefont {Mel'nik}}, \bibinfo {author}
  {\bibfnamefont {N.~J.}\ \bibnamefont {Curro}}, \bibinfo {author}
  {\bibfnamefont {J.~D.}\ \bibnamefont {Thompson}}, \ and\ \bibinfo {author}
  {\bibfnamefont {S.~M.}\ \bibnamefont {Stishov}},\ }\href {\doibase
  10.1038/nature02449} {\bibfield  {journal} {\bibinfo  {journal} {Nature}\
  }\textbf {\bibinfo {volume} {428}},\ \bibinfo {pages} {542} (\bibinfo {year}
  {2004})}\BibitemShut {NoStop}%
\bibitem [{\citenamefont {Takano}\ \emph {et~al.}(2005)\citenamefont {Takano},
  \citenamefont {Nagao}, \citenamefont {Takenouchi}, \citenamefont {Umezawa},
  \citenamefont {Sakaguchi}, \citenamefont {Tachiki},\ and\ \citenamefont
  {Kawarada}}]{takano2005}%
  \BibitemOpen
  \bibfield  {author} {\bibinfo {author} {\bibfnamefont {Y.}~\bibnamefont
  {Takano}}, \bibinfo {author} {\bibfnamefont {M.}~\bibnamefont {Nagao}},
  \bibinfo {author} {\bibfnamefont {T.}~\bibnamefont {Takenouchi}}, \bibinfo
  {author} {\bibfnamefont {H.}~\bibnamefont {Umezawa}}, \bibinfo {author}
  {\bibfnamefont {I.}~\bibnamefont {Sakaguchi}}, \bibinfo {author}
  {\bibfnamefont {M.}~\bibnamefont {Tachiki}}, \ and\ \bibinfo {author}
  {\bibfnamefont {H.}~\bibnamefont {Kawarada}},\ }\href {\doibase
  10.1016/j.diamond.2005.08.014} {\bibfield  {journal} {\bibinfo  {journal}
  {Diamond and Related Materials}\ }\textbf {\bibinfo {volume} {14}},\ \bibinfo
  {pages} {1936} (\bibinfo {year} {2005})}\BibitemShut {NoStop}%
\bibitem [{\citenamefont {Zhang}\ \emph {et~al.}(2017)\citenamefont {Zhang},
  \citenamefont {Samuely}, \citenamefont {Xu}, \citenamefont {Jochum},
  \citenamefont {Volodin}, \citenamefont {Zhou}, \citenamefont {May},
  \citenamefont {Onufriienko}, \citenamefont {Ka{\v{c}}mar{\v{c}}{\'{i}}k},
  \citenamefont {Steele}, \citenamefont {Li}, \citenamefont {Vanacken},
  \citenamefont {Vac{\'{i}}k}, \citenamefont {Szab{\'{o}}}, \citenamefont
  {Yuan}, \citenamefont {Roeffaers}, \citenamefont {Cerbu}, \citenamefont
  {Samuely}, \citenamefont {Hofkens},\ and\ \citenamefont
  {Moshchalkov}}]{zhang2017}%
  \BibitemOpen
  \bibfield  {author} {\bibinfo {author} {\bibfnamefont {G.}~\bibnamefont
  {Zhang}}, \bibinfo {author} {\bibfnamefont {T.}~\bibnamefont {Samuely}},
  \bibinfo {author} {\bibfnamefont {Z.}~\bibnamefont {Xu}}, \bibinfo {author}
  {\bibfnamefont {J.~K.}\ \bibnamefont {Jochum}}, \bibinfo {author}
  {\bibfnamefont {A.}~\bibnamefont {Volodin}}, \bibinfo {author} {\bibfnamefont
  {S.}~\bibnamefont {Zhou}}, \bibinfo {author} {\bibfnamefont {P.~W.}\
  \bibnamefont {May}}, \bibinfo {author} {\bibfnamefont {O.}~\bibnamefont
  {Onufriienko}}, \bibinfo {author} {\bibfnamefont {J.}~\bibnamefont
  {Ka{\v{c}}mar{\v{c}}{\'{i}}k}}, \bibinfo {author} {\bibfnamefont {J.~A.}\
  \bibnamefont {Steele}}, \bibinfo {author} {\bibfnamefont {J.}~\bibnamefont
  {Li}}, \bibinfo {author} {\bibfnamefont {J.}~\bibnamefont {Vanacken}},
  \bibinfo {author} {\bibfnamefont {J.}~\bibnamefont {Vac{\'{i}}k}}, \bibinfo
  {author} {\bibfnamefont {P.}~\bibnamefont {Szab{\'{o}}}}, \bibinfo {author}
  {\bibfnamefont {H.}~\bibnamefont {Yuan}}, \bibinfo {author} {\bibfnamefont
  {M.~B.~J.}\ \bibnamefont {Roeffaers}}, \bibinfo {author} {\bibfnamefont
  {D.}~\bibnamefont {Cerbu}}, \bibinfo {author} {\bibfnamefont
  {P.}~\bibnamefont {Samuely}}, \bibinfo {author} {\bibfnamefont
  {J.}~\bibnamefont {Hofkens}}, \ and\ \bibinfo {author} {\bibfnamefont
  {V.~V.}\ \bibnamefont {Moshchalkov}},\ }\href {\doibase
  10.1021/acsnano.7b01688} {\bibfield  {journal} {\bibinfo  {journal} {ACS
  Nano}\ }\textbf {\bibinfo {volume} {11}},\ \bibinfo {pages} {5358} (\bibinfo
  {year} {2017})}\BibitemShut {NoStop}%
\end{thebibliography}%

\end{document}